\documentclass[aps,prb,twocolumn,groupedaddress,showpacs,showkeys]{revtex4}

\usepackage{xspace}
\usepackage{epsfig}
\usepackage{amsfonts}

\newcommand{\fig}[4]{
\begin{figure}[#1]
\begin{center}
\includegraphics [width=#2 in, keepaspectratio=true] {#3.eps}
\caption{#4 \label{fig:#3}}
\end{center}
\end{figure}}

\begin{document}


\title{Magnetotransport properties of a polarization-doped three-dimensional electron slab}

\author{Debdeep Jena}
\email[Electronic mail: ]{djena@engineering.ucsb.edu}
\author{Sten Heikman}
\author{James S. Speck}
\author{Arthur Gossard}
\author{Umesh K. Mishra}
\affiliation{Department of Electrical and Computer Engineering and Materials Department \\
            University of California, Santa Barbara \\
            CA, 93106}
\author{Angela Link}
\author{Oliver Ambacher}
\affiliation{Walter Schottky Institute, Am Coulombwall 3 D-85748
Garching, Germany }

\date{\today}

\begin{abstract}
We present evidence of strong Shubnikov-de-Haas magnetoresistance
oscillations in a polarization-doped degenerate three-dimensional
electron slab in an Al$_{x}$Ga$_{1-x}$N semiconductor system.  The
degenerate free carriers are generated by a novel technique by
grading a polar alloy semiconductor with spatially changing
polarization. Analysis of the magnetotransport data enables us to
extract an effective mass of $m^{\star}=0.19 m_{0}$ and a quantum
scattering time of $\tau_{q}= 0.3 ps$.  Analysis of scattering
processes helps us extract an alloy scattering parameter for the
Al$_{x}$Ga$_{1-x}$N material system to be $V_{0}=1.8eV$.
\end{abstract}


\maketitle

\section{Introduction}


%


In a recent paper\cite{MyNewAPL}, we reported the creation of
polarization-doped three-dimensional electron slabs (3DES) in a
graded AlGaN layer.  In this work, we present magnetotransport
studies of the polarization-doped 3DES.  The 3DES exhibits strong
Shubnikov de-Haas oscillations.  A study of these oscillations
reveals the effective mass of carriers, the collisional broadening
due to quantum scattering , and the nature of transport in the
polarization-doped 3DES.

In a polar crystal with {\em uniform} composition, $n=\nabla \cdot
{\bf P}=0$ (where ${\bf P}$ is the polarization vector), and there
is no bulk polarization charge.  The polarization of the crystal
manifests only as sheet charges at {\em surfaces} and {\em
interfaces}, owing to the change in polarization across them
($\sigma = ({\bf P}_{1}-{\bf P}_{2}) \cdot {\bf \hat{n}} $, where
$\sigma$ is the sheet charge, ${\bf P_{1}},{\bf P_{2}}$ are the
polarization vectors on the two sides of the junction, and ${\bf
\hat{n}}$ is the normal to the junction).  An alloy of spatially
constant composition of two polar materials with different
magnitudes of polarization will also have $n=\nabla \cdot {\bf
P}=0$.  However, if the composition changes spatially, the
divergence of the polarization vector becomes non-vanishing. This
leads to an {\em immobile} polarization charge $N_{\pi}^{D}=\nabla
\cdot {\bf P}$ in a graded alloy region.

The III-V nitride family of crystals of the wurtzite crystal
structure is among the first material systems that is strongly
polar(compared to other III-V semiconductors) along the c-(0001)
axis, is semiconducting (most strongly polar crystals are
insulators and used as dielectrics), its binaries (GaN, AlN, InN)
have different polarizations, and they can be grown expitaxially
with sufficiently pure crystalline quality. These unique
properties of the material system enables us to exploit the
concept of polarization bulk doping.  The large difference in
polarization between AlN and GaN (both spontaneous and
piezoelectric) \cite{bernardini} and the good control of epitaxial
growth of AlGaN alloys prompted its choice for this experiment.


Figure I shows a schematic of charge control and band diagram of
the technique of polarization doping that we have employed.  Also
shown in the figure is the sample structure we have used. The
sample is a Ga-face structure grown by plasma-induced molecular
beam epitaxy \cite{ben} on a metal-organic chemical vapor
deposition grown semi-insulating\cite{sten} GaN on a sapphire
substrate.  The growth is along the polar c(0001)
axis\cite{patrick}.  The top 100nm of the structure is linearly
graded AlGaN; the composition of Al is changed from 0-30\% by
controlling the aluminum flux by a computer
program\cite{MyNewAPL}.  This is the layer that is polarization
doped.

The linear grading leads to a fixed charge\cite{notenonlinear}
doping $N_{\pi}^{D}=\nabla \cdot {\bf P}$ terminated by a opposite
sheet charge $\sigma_{\pi}^{S}=|{\bf P}|/\epsilon$ at the surface.
The large electric field created by the jellium of fixed
polarization charge extracts electrons from surface donor states
\cite{ibbo} to form a mobile 3-dimensional electron slab (3DES).
The mobile 3DES is characterized by a temperature independent
carrier concentration for $0.4K \leq T \leq 300K$ verifying that
the carriers are degenerate, and thus distinct from carriers
activated from shallow donors \cite{MyNewAPL}.  We now proceed to
study the magnetotransport properties of the 3DES created by this
novel process.


\section{Magnetotransport}

In the presence of a quantizing magnetic field, the unperturbed
3-dimensional density of states (DOS) $g_{0}(\varepsilon)$
undergoes Landau quantization to quasi 1-dimensional density of
states and acquires an oscillatory component.  Dingle\cite{dingle}
showed that inclusion of collisional broadening removes the
divergence at the bottom of each 1-dimensional subband and damps
the DOS oscillation amplitudes exponentially in $1/B$.

\fig{t}{2.9}{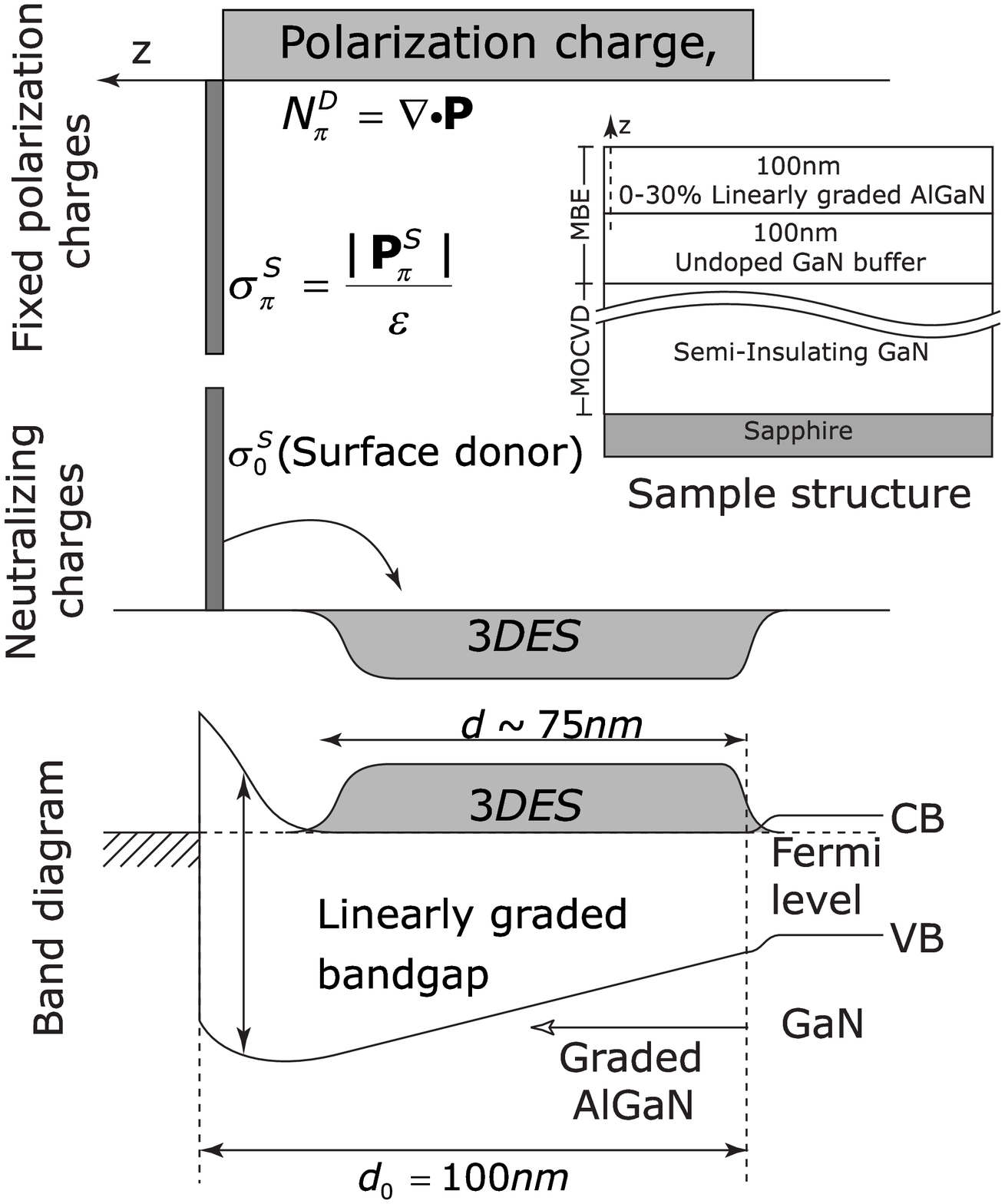}{Schematic of charge control
showing polarization charges and formation of the 3DES.  The band
diagram shows depletion of the 3DES from the surface potential.
Also shown is the epitaxial layer structure that generates the
3DES.}

As is well known from the theory of the magnetic quantum effects,
this oscillation of the density of states manifests in
oscillations of both the diamagnetic susceptibility (manifesting
in the de-Haas Van Alphen effect) and transport coefficients
(manifesting in the Shubnikov de-Haas or SdH oscillations).  In
particular, the transverse (${\bf B} \perp {\bf E}$)
magnetoresistance $R_{xx}$ shows oscillations in $1/B$.  Kubo
\cite{kubo} derived the transverse magnetoresistance at high
magnetic fields using the density-matrix approach to solve the
transport problem.  The expression for $R_{xx}$ can be decomposed
into a background part and an oscillatory
contribution\cite{rothargyres} $R_{xx}=R_{xx}^{Back}+ \Delta
R_{xx}^{osc}$.  The background term is attributed to sample
inhomogeneities and disorder.  The amplitude ($A$) of the
oscillatory component can be cast in a form \cite{hamaguchi} that
is simple to use and captures the physical processes reflected in
the measured magnetoresistance -

\begin{equation}
\Delta R_{xx}^{osc} \propto A = \underbrace{ \frac{\chi}{\sinh
\chi}}_{D_{t}(T)} \times \underbrace{\exp(-\frac{2 \pi
\Gamma}{\hbar \omega_{C}})}_{D_{c}(B)} \times  (\frac{\hbar
\omega_{c}}{2 \varepsilon_{F}})^{1/2} \cos(\frac{2 \pi
\varepsilon_{F}}{\hbar \omega_{C}} - \delta )
\end{equation}

where $\Delta R_{xx}^{osc}$ is the oscillating part of the
magnetoresistance with the background removed,
$\varepsilon_{F}=(\hbar^{2}/2 m^{\star})(3 \pi^{2}n_{3DES})^{2/3}$
is the Fermi energy of the 3DES, $\hbar$ being the reduced
Planck's constant and $n_{3DES}$ is the carrier density of the
3DES, and $\omega_{C}=eB/m^{\star}$ is the cyclotron frequency.
$\chi=2 \pi^{2}k_{B}T/\hbar \omega_{c}$ is a temperature dependent
dimensionless parameter and $\Gamma$ is the collisional broadening
energy due to quantum scattering events.  $\delta$ is a phase
factor that is unimportant for our study.  The terms $D_{t}(T)$
and $D_{c}(B)$ are the temperature and collision damping terms
respectively; it is easily seen that in the absence of damping of
the oscillations due to temperature ($\lim_{T \rightarrow 0}
D_{t}(T) = 1$) and in the absence of damping due to collisions
($\lim_{\Gamma \rightarrow 0} D_{c}(B) = 1$), the
magnetoresistance would exhibit a weakly modulated ($\sim
B^{1/2}$) cosine oscillations in $1/B$. In fact, the two damping
terms $D_{t}(T),D_{c}(B)$ are used as probes to tune the
temperature and magnetic field independently to extract the
effective mass and the quantum scattering time.  The period of the
cosine oscillatory term yields the carrier density of the 3DES
since the period is linked to $n_{3DES}$.  $R_{xy}$, the Hall
resistance is linear with $B$, and should show plateaus at the
minima of $R_{xx}$ when a small number of Landau levels are
filled.

\fig{t!}{2.9}{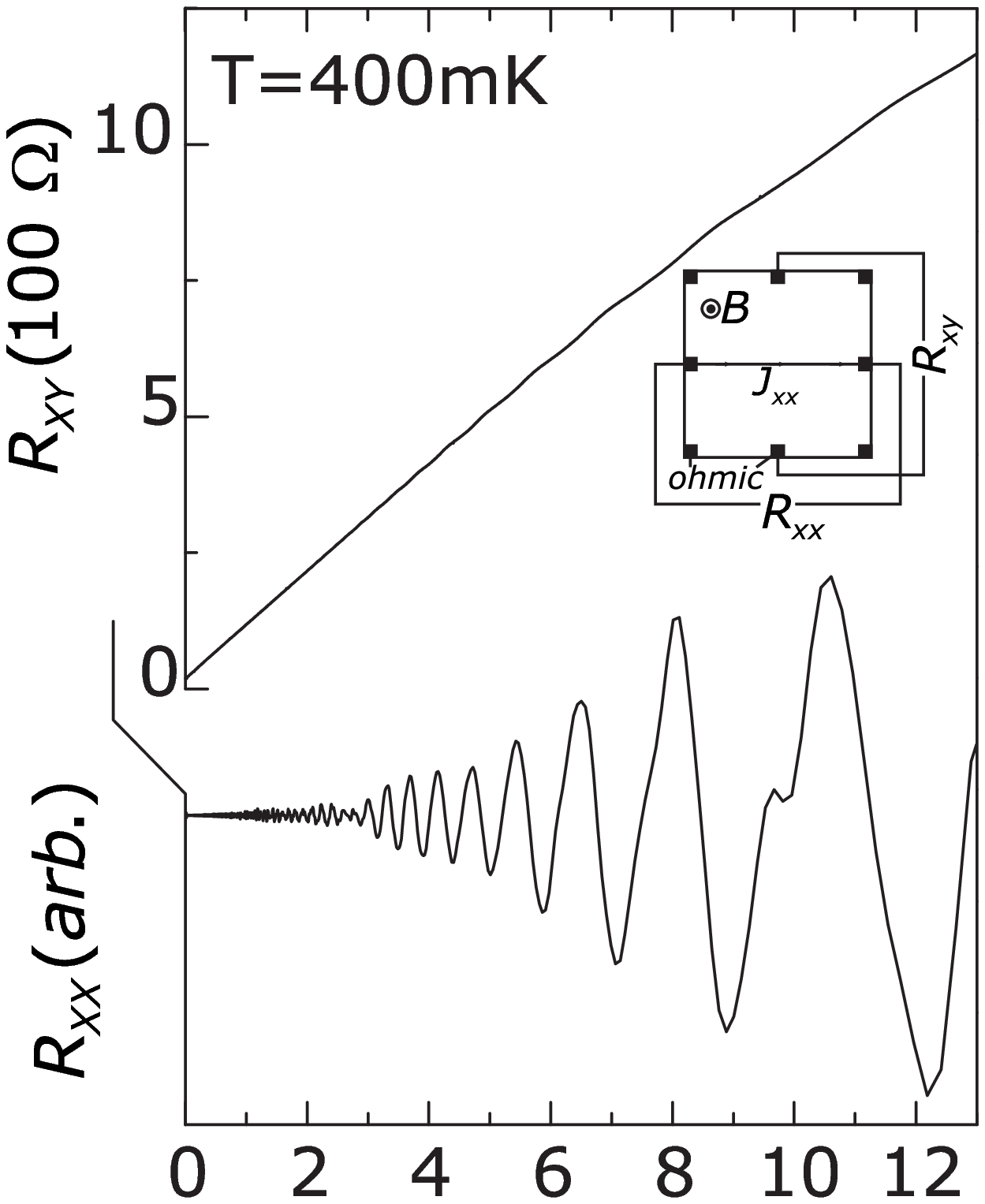}{Magnetotransport measurement
data at $T=400mK$.  The figure also shown as an inset the geometry
used for measuring $R_{xx},R_{xy}$.  The $R_{xx}$ shown is the
oscillatory component with the background removed.}


For magnetotransport measurements on our 3DES, ohmic contacts were
formed in a Van-der Pauw geometry (Figure 1 inset).  The sample
was immersed in a $^{3}He$ low-temperature cryostat with a base
temperature of $300mK$.  Magnetic fields in the range $0T \leq B
\leq 14T$ were applied. $R_{xx}$ and $R_{xy}$ was measured as in
the geometry depicted in the figure using the standard
low-frequency lock-in technique.

In Figure II we show a plot of the oscillatory transverse
magnetoresistance $\Delta R_{xx}^{osc}$ and $R_{xy}$ at $T=400mK$
plotted against the applied magnetic field.  We have removed the
background using a FFT filter for plotting $R_{xx}$. We will
briefly describe our FFT process later in this work.  The Hall
mobility determined from the slope of the $R_{xy}$ curve is
$\mu_{H} \simeq 3000cm^{2}/V \cdot s$, which is higher than 77K
low-field Hall mobility of $\mu_{77K} \simeq 2500cm^{2}/V \cdot
s$.  Also, assuming that the 3DES is spread over a thickness $d$,
the sheet carrier density of the 3DES is calculated to be
$n_{3DES}\times d =1/R_{H}e=B/eR_{xy}=7.2 \times 10^{12}/cm^{2}$.
This is consistent with the 77K low-field Hall measured sheet
density of $7.5 \times 10^{12}/cm^{2}$.  The spread of the 3DES is
calculated from a self-consistent Poisson-Schrodinger band
calculation to be $d=75nm$ due to of $25nm$ depletion of the 3DES
from the surface potential.  This depletion in the graded AlGaN
layer has also been verified by capacitance-voltage
profiling\cite{MyNewAPL}.  Thus, the Hall 3-dimensional carrier
density is $n_{3DES} \sim 10^{18}/cm^{3}$.


\section{Analysis of magnetotransport data}

We begin our study by analyzing the oscillatory component of the
transverse magnetoresistance (Figure III).  For achieving this, we
first take the raw $R_{xx}$ vs $B$ data and interpolate it to
create an equally spaced $N=2^{15}$ size FFT window.  We then find
the FFT power spectrum.  This is repeated for $R_{xx}$ measured at
different temperatures.  A typical FFT power spectrum (at T=2.5K)
is shown in the inset of Figure III. There is a clearly resolved
peak at the fundamental oscillation period $B_{0}=34.01T$.  A band
pass filter [$f_{pass}=$28-150T] is then employed to remove the
background component.  The resulting $\Delta R_{xx}^{osc}$ for
various temperatures $0.4 K < T < 9.5 K$ is plotted against $1/B$
in Figure III.  As is clear from the plot, the period of
oscillations is $\Delta(\frac{1}{B})=0.0294T^{-1}=1/B_{0}$.  The
oscillations are strongly damped with increasing $1/B$ as well as
with increasing temperature, as predicted by the theory (Equation
1).


\subsection{Carrier concentration}

First, we observe from Equation 1 that the period
$\Delta(\frac{1}{B})$ is linked to the carrier density of the 3DEG
by the relation $\Delta (1/B) = e\hbar/ m^{\star}\varepsilon_{F} =
\frac{2e}{\hbar}(3 \pi^{2}n_{3DES})^{-2/3} $.  From the plot, the
period $\Delta (1/B) = 0.0294$T$^{-1}$ yields a direct measurement
of the 3-dimensional carrier concentration $n_{3D}^{SdH}=1.1
\times 10^{18}cm^{-3}$.  Thus, the carrier density measured from
the quantum oscillations is close to the carrier density measured
by classical Hall technique ($n_{3DES}=10^{18}cm^{-3}$).

\fig{t!}{3.2}{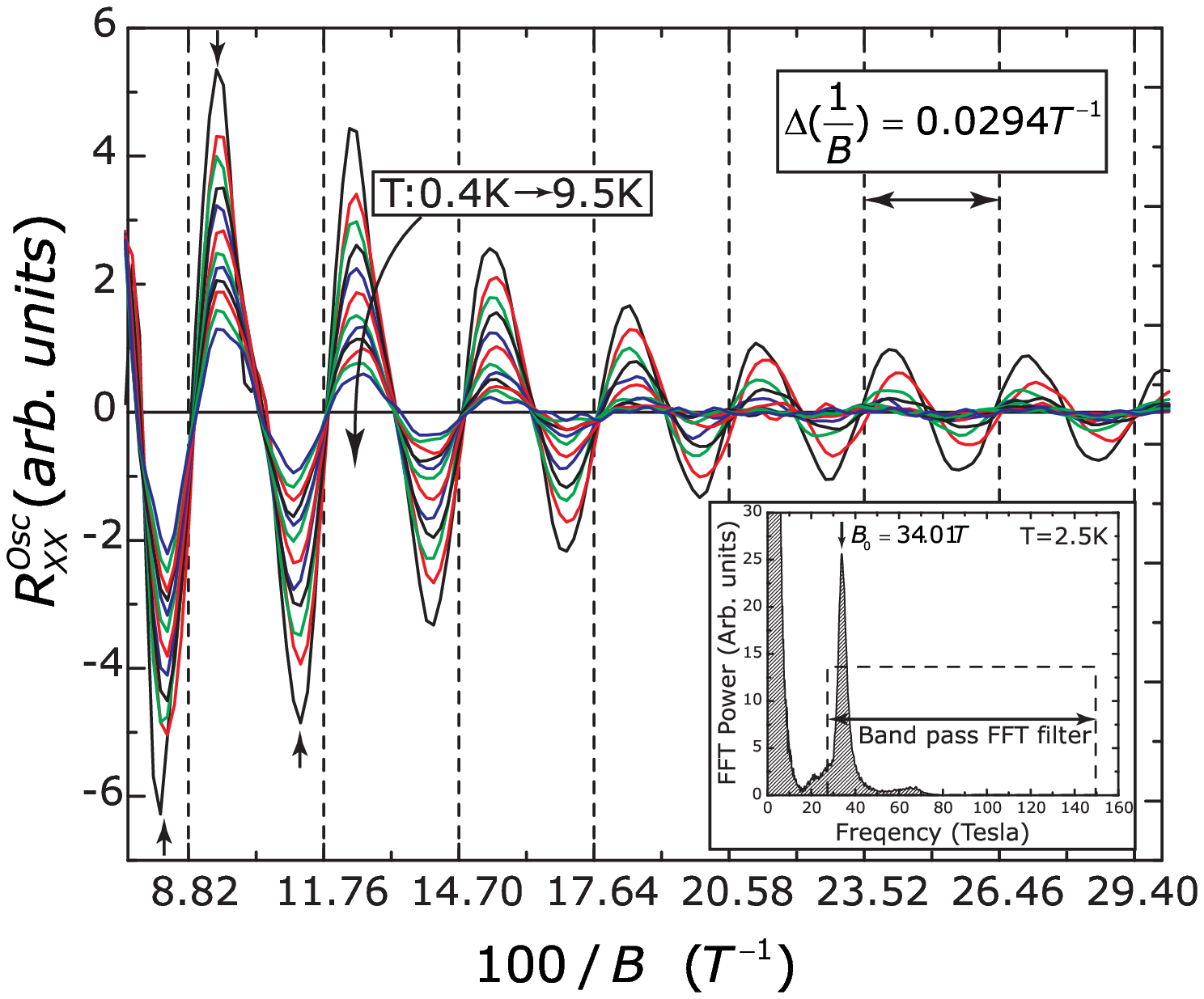}{The oscillatory component $\Delta
R_{xx}^{osc}$ plotted against $1/B$. The oscillations are periodic
with period $\Delta(1/B)=0.0294T^{-1}$, and are damped with both
increasing temperature (different curves), and increasing $1/B$.
Also shown in the inset is a typical FFT power spectrum (at
$T=2.5K$) showing a peak at the fundamental period, and the
band-pass window used to filter the oscillatory component $\Delta
R_{xx}^{osc} $.}


\subsection{Effective mass}

The second important property of the 3DES that is accessible from
the SdH oscllations is the effective mass of the electrons.  This
can be calculated from a controlled temperature damping of the
oscillation amplitudes, i.e., by tuning the term $D_{t}(T)$ (which
is the only temperature dependent term).  The
approximation\cite{elhamri} $\sinh(\chi) \sim exp(\chi)/2$
converts Equation 1 into

\begin{equation}
\ln(\frac{A}{T}) = C - (\frac{2 \pi^{2}k_{B}m^{\star}}{e \hbar
B})T
\end{equation}

where the constant $C$ absorbs all terms independent of
temperature.  Thus, when $\ln(\frac{A}{T})$ is plotted against $T$
for a fixed magnetic field, it results in a straight line with
slope $S=-\frac{2 \pi^{2}k_{B}m^{\star}}{e \hbar B}$.  Since the
effective mass is the only unknown, it can be calculated.  We
calculate the effective mass by employing this procedure.

We choose the amplitude maxima at three different magnetic fields
$B=8.9T,10.5T,12.0T$ for extracting effective mass of the 3DES.
These maxima are indicated by arrows in Figure 3.  The plots of
$\ln(A/T)$ against temperature ($2.5K<T<14.5K$) for these magnetic
fields are shown in Figure IV(a).  The slopes of the plots at
$B=8.9T,10.5T,12.0T$ yield effective masses
$m^{\star}=0.189,0.197,0.189m_{0}$ respectively.  The accepted
effective mass of electrons in bulk GaN is $m^{\star}=0.2m_{0}$.
SdH measurement of electron effective mass in bulk GaN is
difficult since the activated carriers in donor-doped samples
freeze out at low temperatures \cite{MyNewAPL}.  Since the
electrons in our 3DES are degenerate, they resist freezeout
effects and remain independent of temperature.  However, the 3DES
electron is not in bulk GaN but in an alloy with a changing
composition; we expect the effective mass to be a spatial average.
The effective mass of electrons in AlN is
predicted\cite{vurgaftman} to be $m^{\star}=0.32m_{0}$; a linear
interpolation gives an expected effective mass of
$m^{\star}=0.21m_{0}$ at an average alloy composition of $ \langle
x \rangle = (0.75 \times 0.3)/2 = 0.11 $ for the mobile 3DES. This
is slightly higher than our measured average value
$m^{\star}_{av}=0.19m_{0}$.  The result is reasonable within
limits of experimental error.  We also note that the effective
mass of electrons in two-dimensional electron gases (2DEGs) at
AlGaN/GaN heterojunctions has been measured by SdH oscillation
technique \cite{elhamri}, \cite{knap}, \cite{saxler},
\cite{arakawa} and reported to be various values around
$m^{\star}\sim 0.2m_{0}$, which is close to our measured value.

\subsection{Scattering times}

The third important parameter of the 3DES that is measured from
the SdH oscillations is the collisional broadening energy $\Gamma$
(due to Dingle).  This term is a measure of the smearing of the
delta-function discontinuities in the DOS due to quantum
scattering events, and it appears as the imaginary part of the
single-particle self energy function.  Collisional broadening
energy is linked to the quantum scattering time $\tau_{q}$ and the
Dingle temperature $T_{D}$ by the relation
$\Gamma=\hbar/2\tau_{q}=\pi k_{B}T_{D}$. This quantity is
experimentally accessible from a controlled Landau damping of the
oscillation amplitudes with $1/B$ at a fixed temperature; in other
words, by tuning $D_{c}(B)$. Equation 1 can be cast in the form

\begin{equation}
\ln(\frac{A^{\star}}{(\frac{\hbar \omega_{c}}{2
\varepsilon_{F}})^{\frac{1}{2}}D_{t}(T)}) = C - (\frac{\pi
m^{\star}}{e \tau_{Q}})\frac{1}{B}
\end{equation}

for extracting $\tau_{q}$ and the related quantities $\Gamma ,
T_{D}$.  Here $A^{\star}$ are the extrema points of the damped
oscillations, forming the exponentially decaying envelope.
Equation 3 suggests that a plot of the natural log of the left
side quantity against $1/B$ (`Dingle plot')should result in a
straight line whose slope is $-\pi m^{\star}/e\tau_{q}$. Since we
have already measured the effective mass $m^{\star}$, we can
extract $\tau_{q}$ from the slope.

\fig{t!}{3.2}{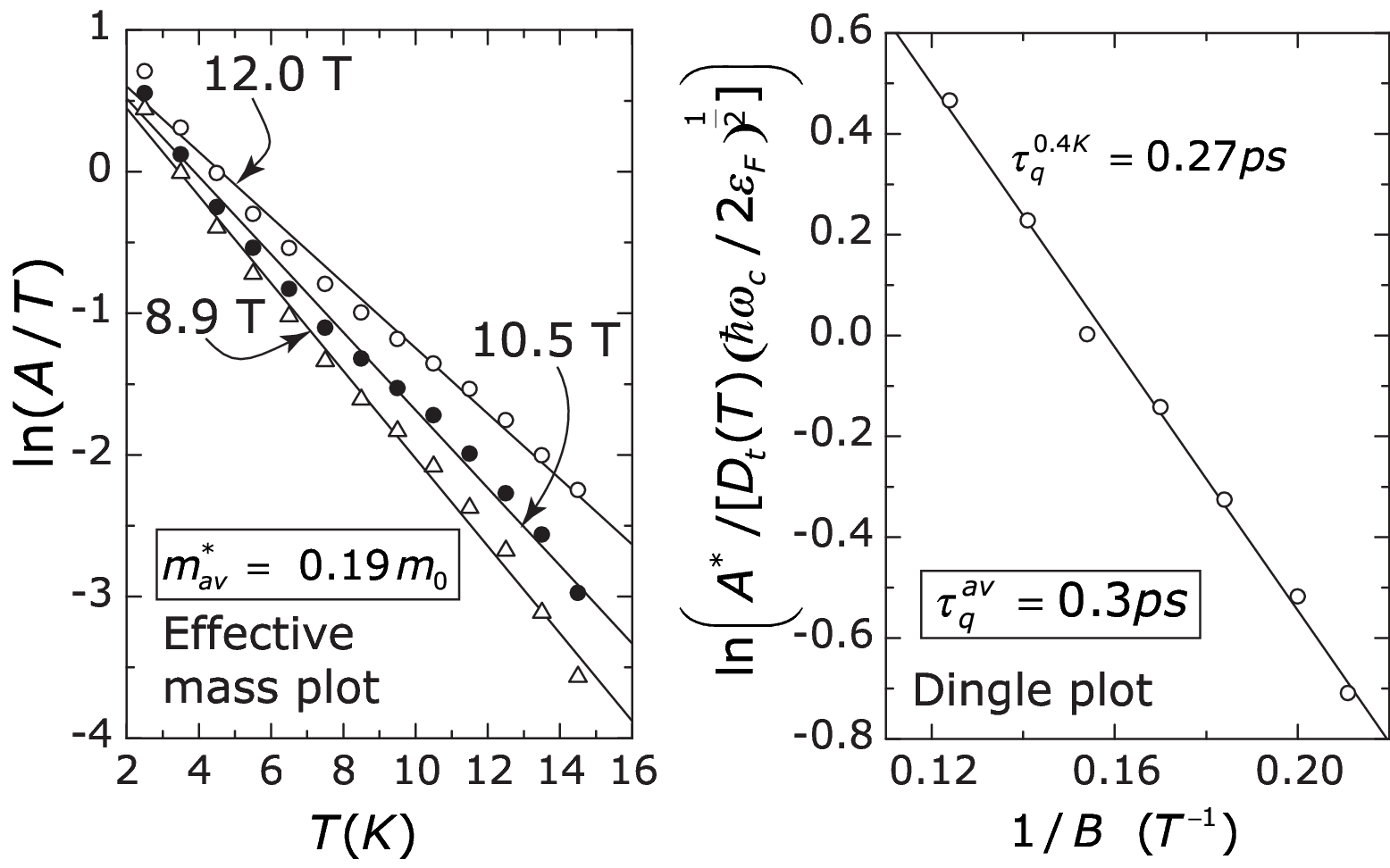}{Effective mass plot and Dingle plot of
the 3DES.  For effective mass, $ln(A/T)$ is plotted for three
values of magnetic fields ($B=8.9T,10.5T,12.0T$) against
temperature.  The slope yields the effective masses
$0.189m_{0},0.197m_{0},0.189m_{0}$ respectively.  The Dingle plot
yields a quantum scattering time of $\tau_{q}=0.27ps$ at $T=400mK$
and an average value of $\tau_{q}^{av}=0.3ps$.}

Figure IV(b) shows the Dingle plot for $T=400mK$, yielding a
$\tau_{q}=0.27ps$.  An averaging of the quantum scattering times
over a range of low temperatures yields a value
$\tau_{q}^{av}=0.3ps$.  The quantum scattering time does not show
any discernible trend with temperature in this range.  We
calculate the corresponding level broadening $\Gamma=1.1meV$ and
Dingle temperature $T_{D}=4K$.  We mention in passing that the
Landau level separation at $B = 10T$ is $\hbar \omega_{C}=5.8meV$,
sufficiently larger than both the measured collisional broadening
of the Landau levels ($\Gamma=1.1meV$) and the thermal broadening
$k_{B}T = 0.09meV$ at $T=1K$, thus satisfying the conditions
required for clear Shubnikov de-Haas oscillations.


Whilst $\tau_{q}$ (the quantum lifetime of an electron in a
magnetic quantum state) is determined by {\em all} scattering
events, the transport (classical, or momentum) lifetime $\tau_{c}$
is weighted by a scattering angle factor $(1-\cos\theta)$, where
$\theta$ is the angular deviation in the scattering event.  Thus,

\begin{equation}
\frac{1}{\tau_{q}}= \int P(\theta) d\theta
\end{equation}

as opposed to the momentum scattering time, which is averaged by
the angular contribution for small angle scattering

\begin{equation}
\frac{1}{\tau_{c}}= \int P(\theta)(1-\cos\theta) d\theta
\end{equation}

here $P(\theta)$ is the transition probability between the initial
and the scattered states determined by Fermi's golden rule.  It
can be seen that for isotropic scattering events with no angular
preference ($P(\theta)$ is independent of $\theta$), the quantum
and classical scattering times are the same $\tau_{c}/\tau_{q}=1$.
If the dominant scattering process has a strong angle dependence,
the ratio is much larger than unity.  This fact has been utilized
to identify the dominant scattering mechanism in modulation-doped
AlGaAs/GaAs two-dimensional electron gases \cite{harrang}.

The low-temperature Hall mobility gives us a direct measurement of
the classical scattering time for the 3DES via the relation
$\mu_{H}=e\tau_{c}/m^{\star}$; the value is $\tau_{c}=0.34ps$.
Within limits of experimental error, the ratio $\tau_{c}/\tau_{q}
\sim 1 $, i.e., is close to unity.

\subsection{Scattering mechanisms}

As Hsu and Walukiewicz have shown\cite{hsu_walu}, one has to
exercise caution before declaring that the dominant scattering
process in the 3DES is of short-range nature.  They show that it
is possible to have the ratio close to unity if there are {\em
different} scattering mechanisms responsible for the quantum and
classical lifetimes.  At the low temperatures at which we measure
the scattering time, both acoustic and optical phonon modes of
scattering are frozen out and the charges from any unintentional
background donors are frozen onto the donor sites, rendering them
neutral.  Size effect scattering \cite{walu_size} that occurs if
the width of the 3DES is much less than the mean-free path of
electrons is negligible since our 3DES has a mean free path
$\lambda=\hbar k_{F} \mu / e \approx 60 nm $ whereas the width of
the 3DES is $d_{0}\approx 75nm$.  The chief scattering mechanisms
that can affect mobility are alloy disorder scattering (since the
3DES is in a graded alloy), charged dislocation scattering (owing
to the high density of dislocations $N_{disl}\approx
10^{9}cm^{-2}$), and ionized impurity scattering from the remote
donors at the surface states.

Hsu and Walukiewicz \cite{hsu_walu} show that remote ionized
impurity scattering strongly favors small angle scattering, thus
causing the ratio $\tau_{c}/\tau_{q} \gg 1$. Since
$\tau_{c}/\tau_{q} \approx 1$ for our 3DES, remote ionized
impurity scattering is unimportant.

The ratio of classical to quantum scattering times due to charged
dislocation scattering has not been found yet. Look\cite{look_prl}
recently applied the classical scattering time due to charged
dislocations and applied it to explain the mobility of bulk doped
GaN samples.  The ratio classical to quantum scattering times due
to charged dislocation scattering in a degenerate 3DES is given
by\cite{mynewpaper}

\begin{equation}
\frac{\tau_{disl}^{c}}{\tau_{disl}^{q}}=1+2k_{F}^{2}\lambda_{TF}^{2}
\end{equation}

where $k_{F}=(3\pi^{2}n_{3DES})^{1/3}$ is the Fermi wavevector and
$\lambda_{TF}^{2}=2 \epsilon \varepsilon_{F} / 3 e^{2} n_{3DES}$
is the Thomas-Fermi screening length for a degenerate 3DES.  The
ratio, which depends only on the 3D carrier density for a
degenerate 3DES, is evidently greater than unity; for our 3DES,
the ratio is $2.3$, which is larger than what we observe.  So
dislocation scattering is not the dominant scattering mechanism
for determining the quantum scattering rate.  The dislocation
scattering dominated classical scattering rates is strongly
dependent on the amount of charge in the dislocation cores, which
is low for our 3DES due to the lack of shallow
dopants\cite{wright}. Thus, we exclude dislocation scattering to
be the most important scattering mechanism for classical
scattering rate.

So we converge on alloy scattering as the dominant scattering
mechanism for low temperatures.  The quantum and classical
scattering rates for alloy disorder scattering are the same since
the scattering potential $V_{0}$ is of a short range nature, which
makes the scattering process {\em angle independent}.

The scattering rate due to alloy disorder with a short range
potential $V_{0}$ has been shown\cite{hamaguchi} to be given by

\begin{equation}
\frac{1}{\tau_{alloy}(k)}=\frac{2\pi}{\hbar} V_{0}^{2}
\Omega(x)x(1-x) g_{3D}(\varepsilon_{k})
\end{equation}

where $\Omega_{0}(x)$ is alloy composition-dependent volume of the
unit cell over which the alloy scattering potential $V_{0}$ is
effective, and $x$ is the alloy composition. $g_{3D}(\varepsilon)$
is the 3-dimensional DOS.  Since the 3DES is highly degenerate,
only carriers with wavevectors very close to $k=k_{F}$ contribute
to transport at low temperatures; hence, the scattering rate is
evaluated at the Fermi energy
$\tau_{alloy}^{-1}=\tau_{alloy}^{-1}(\varepsilon_{F})$. Besides,
the alloy is graded, and Matheissen's rule which is accurate for
low temperature transport analysis is used for a spatial averaging
of the scattering rate

\begin{equation}
\langle \tau_{alloy}^{-1} \rangle = \frac{1}{x_{0}}
\int_{0}^{x_{0}} \tau_{alloy}^{-1}(x)dx
\end{equation}

where $x_{0}=0.225$ is the alloy composition experienced by 3DES
electrons at the edge of the depletion region.  Using this simple
result, we conclude that to achieve a low-temperature transport
mobility of $3000cm^{2}/V \cdot s$, an alloy scattering potential
of $V_{0}=1.8eV$ is necessary.

It is important to point out that the alloy scattering potential
as defined in the model for alloy disorder is at best a fitting
parameter.  Due to the lack of experimental values, it is common
practice to assume the scattering potential to be the conduction
band offset between the binaries forming the alloy ($V_{0}=\Delta
E_{c}=2.1eV$ for AlN,GaN).  We note that with an alloy scattering
potential of $V_{0}=2.1eV$, the calculated mobility is lower
($\approx 2000cm^{2}/V \cdot s$) than the measured value. Besides,
our 3DES mobility is dominated by alloy scattering and all other
scattering mechanisms are removed by measuring the mobility at low
temperatures, making it a clean measurement of the alloy
scattering potential.  Our report presents the first direct
measurement of the alloy scattering potential in the AlGaN
material system.


\section{Summary}

We finally summarize our findings in this work (see Table I).  By
exploiting the polarization charges in the AlGaN/GaN semiconductor
system, we are able to create a 3DES without intentional doping.
The mobile carriers in the 3DES are degenerate and exhibit a high
mobility even to the lowest temperatures.  The lack of carrier
freezeout enables us to observe Shubnikov de-Haas oscillations in
magnetotransport measurements of the 3DES.  The oscillations
reveal several important facts about the 3DES.  First, the
temperature damping of oscillations reveals the effective mass of
electrons to be very close to that in bulk GaN
($m^{\star}=0.19m_{0}$).  Next, the quantum scattering time of
electrons in the 3DES is found from the Dingle plot to be
$\tau_{q}=0.3ps$.  The ratio of the classical (momentum)
scattering time to the classical (momentum) scattering time is
found to be close to unity $\tau_{c}/\tau_{q} \approx 1$. The
ratio suggests predominantly short-range scattering dominating
transport properties at low temperatures.  This scattering
mechanism is identified to be alloy scattering.  This lets us
extract another valuable parameter, the alloy scattering potential
in Al$_{x}$Ga$_{1-x}$N to be $V_{0}=1.8eV$.

\begin{acknowledgments}

The authors would like to thank Jasprit Singh, and Herbert Kroemer
for useful discussions, and Huili Xing for a critical reading of
the manuscript.  Funding from POLARIS/MURI (Contract monitor: C.
Wood) is gratefully acknowledged.

\end{acknowledgments}

\bibliographystyle{ieeetr}
\bibliography{../MainRefs}

\begin{references}
\bibitem{notenonlinear}  There is a small nonlinearity associated with the piezoelectric component which can be neglected.
\end{references}

\pagebreak


\begin{table*}[t!]
\caption{Summary of constants used and results extracted from
magnetoresistance measurements} \label{tab:samples}
\begin{tabular}{c c c c}
\hline\hline
Quantity                    & Symbol        & Magnitude             & Unit \\
\hline

Relative dielectric constant& $\epsilon_{r}$& 8.9(GaN), 8.5(AlN) & - \\
Lattice constant            & $a_{0}$       & 3.189(GaN), 3.112(AlN) & \AA \\
Lattice constant            & $c_{0}$       & 5.185(GaN), 4.982(AlN) & \AA \\

\hline
Effective mass              & $m^{\star}$   & 0.19                  & $m_{0}$\\
Quantum scattering time     & $\tau_{q}$    & 0.3                   & ps  \\
Transport scattering time   & $\tau_{m}$    & 0.34                  & ps  \\
3DES density                & $n_{3DES}$    & 1.1 $\times$ 10$^{18}$   & $cm^{-3}$ \\
3DES Hall mobility (1K)     & $\mu_{H}$     & 3000                  & $cm^{2}/V \cdot s$ \\
Alloy scattering potential  & $V_{0}$       & 1.8                   & $eV$ \\

\hline\hline
\end{tabular}
\end{table*}

\end{document}